\documentclass[prl,aps,showpacs,twocolumn,preprintnumbers,
amsmath,amssymb,superscriptaddress,longbibliography]{revtex4-1}
\usepackage[english]{babel}
\usepackage{amsmath,amssymb,amsfonts}
\usepackage{graphicx}
\usepackage[colorlinks=True,linkcolor=red,citecolor=blue,urlcolor=blue]{hyperref}
\usepackage{bookmark}
\usepackage{xcolor}
\usepackage{braket}
\usepackage{nicefrac}
\usepackage{bm}
\usepackage{bbm}
\usepackage{braket}
\usepackage{slashed}
\usepackage[normalem]{ulem}
\usepackage{array}
\newcolumntype{C}{>{$}c<{$}}

\renewcommand{\Im}{\mathop{\mathrm{Im}}}

\newcommand{\ave}[1]{\langle #1 \rangle}
\newcommand{\bolds}[1]{\boldsymbol #1}

\allowdisplaybreaks

  \DeclareUnicodeCharacter{2212}{-}

\makeatletter
\newcommand\thankssymb[1]{\lowercase{\textsuperscript{\@alph{#1}}}}
\makeatother

\begin{document}

\title{Supplemental Material: "Nematicity-enhanced superconductivity in systems with a non-Fermi liquid behavior"}

\author{Sharareh Sayyad}
\email{sharareh.sayyad@neel.cnrs.fr}
\affiliation{Univ. Grenoble Alpes, CNRS, Grenoble INP, Institut N\'eel, Grenoble 38000, France}
\affiliation{Max Planck Institute for the Science of Light, Staudtstra\ss e 2, 91058 Erlangen, Germany}

\author{Motoharu Kitatani \normalfont\textsuperscript{a,}}
\affiliation{RIKEN Center for Emergent Matter Sciences~(CEMS), Wako, Saitama, 351-0198, Japan}
\thanks{a. These two authors contributed equally}

\author{Abolhassan Vaezi \normalfont\textsuperscript{a,}}
\email{vaezi@sharif.edu}
\affiliation{Department of Physics, Sharif University of Technology, Tehran 14588-89694, Iran}
\thanks{a. These two authors contributed equally}

\author{Hideo Aoki}
\affiliation{Department of Physics, University of Tokyo, Hongo, Tokyo 113-0033, Japan}
\affiliation{Electronics and Photonics Research Institute, Advanced Industrial Science and Technology~(AIST), Tsukuba, Ibaraki 305-8568, Japan}

\begin{abstract}

We explore the interplay between nematicity~(spontaneous breaking of the sixfold rotational symmetry), superconductivity, and non-Fermi liquid behavior in partially flat-band models on the triangular lattice.
A key result is that the nematicity (Pomeranchuk instability), which is driven by many-body effect and stronger in flat-band systems, enhances superconducting transition temperature in a systematic manner on the $T_{\rm c}$ dome. There, a plausible $s_{x^2+y^2} -d_{x^2-y^2} - d_{xy}$-wave symmetry, in place of the conventional $d_{x^2-y^2}$-wave, governs the nematicity-enhanced pairing with a sharp rise in the $T_{\rm c}$ dome on the filling axis. 
When the sixfold symmetry is spontaneously broken, the pairing interaction is shown to become stronger with more compact 
pairs in real space than when the symmetry is enforced.
These are accompanied by a non-Fermi character of electrons in the partially flat bands with many-body interactions. 

\end{abstract}

\maketitle

\section{ Introduction}

Strongly correlated systems have become an epitome in the condensed-matter physics, as exemplified by the high-temperature superconductivity in the cuprate~\cite{ Keimer2014}, and iron-based~~\cite{ Hosono2015} families. These compounds exhibit rich phase diagrams as hallmarked by the emergence of unconventional superconductivity, 
 and a plethora of symmetry-broken phases such as spin and charge nematicity and stripe orders. 

Quest for finding novel high-temperature superconductors spurs interests in exploring many-body systems with short-range repulsions but with~(nearly) flat subregions in the band dispersion arising from hopping beyond nearest neighbors or from lattice structures~\cite{Nogueira1996, Aoki2019}. These systems with dispersionless band portions permit numerous scattering channels for the electrons and can give rise to various exotic quantum phases such as spin and charge density waves~\cite{Kozii2019, Fernandes2020}, Mott insulating~\cite{Chen2020}, and bad-metallic phases~\cite{Huang2019}, as well as the formation of spatially extended Cooper pairs~\cite{Sayyad2020, Yuan2019}.

Interaction and the flatness of the band structure can be intimately related to geometric and quantum frustration in producing strong correlation effects. The spin liquid behavior in hexagonal lattices, such as organic compounds~\cite{Itou2010, Yamashita2011}) and inorganic Herbertsmithites~\cite{Norman2016, Zhong2019}, are typical examples. In these exotic liquids, the classical picture is no longer valid, and their quantum phase transitions cannot be described within Landau's phase transition theory.

Aside from these many-body phenomena, the electron nematicity, i.e., spontaneous breaking of spatial rotational symmetry triggered by many-body interactions, is another manifestation of the correlation effects~\cite{Halboth2000, Yamase2000}. It is an intriguing direction to pursue the physical origins of these symmetry-broken phases~\cite{Fedorov2016, Gallais2016, Lee2018}, and to grasp the interplay between nematicity and other phases such as superconductivity~\cite{Fradkin2010, Venderbos2018,  Dodaro2018, Yuan2019, Chen2020b} and non-Fermi liquid~\cite{Edalati2012}. Various studies report different roles of nematic fluctuations on the superconductivity including the competition between these two phases, e.g., in doped $\rm BaFe_{2}As_{2}$~\cite{Malinowski2020}, the assistance of nematicity to enhance the superconductivity transition temperature, e.g., in twisted bilayer graphene~\cite{Cao2021}, and the negligible effect of nematicity on the superconducting phase, e.g., in $\rm FeSe$~\cite{Bohmer2013}. One crucial aspect is figuring out which of these possibilities occur in systems that have a flat or partially flat band in their dispersions~\cite{Kozii2019, Isobe2018, Fernandes2019, Fernandes2020}.

In this paper, we bring these features together to explore the interplay between the nematicity and superconductivity in partially flat band~(PFB) models on the triangular lattice, effective model for Moire-produced basis states~\cite{Biderang2022}. Here the lattice structure frustrates magnetic orders, thereby giving opportunities for nematic instabilities to arise. 
As a key finding, we shall demonstrate that nematicity can significantly enhance transition temperatures~($T_{\rm c}$) in the superconducting phase, with a $s_{x^2+y^2} -d_{x^2-y^2}-d_{xy}$-wave pairing symmetry. This occurs for an intermediate Hubbard repulsion and in a non-Fermi liquid regime.

   \begin{figure}[t]
    \centering
    \includegraphics[width=0.9\columnwidth]{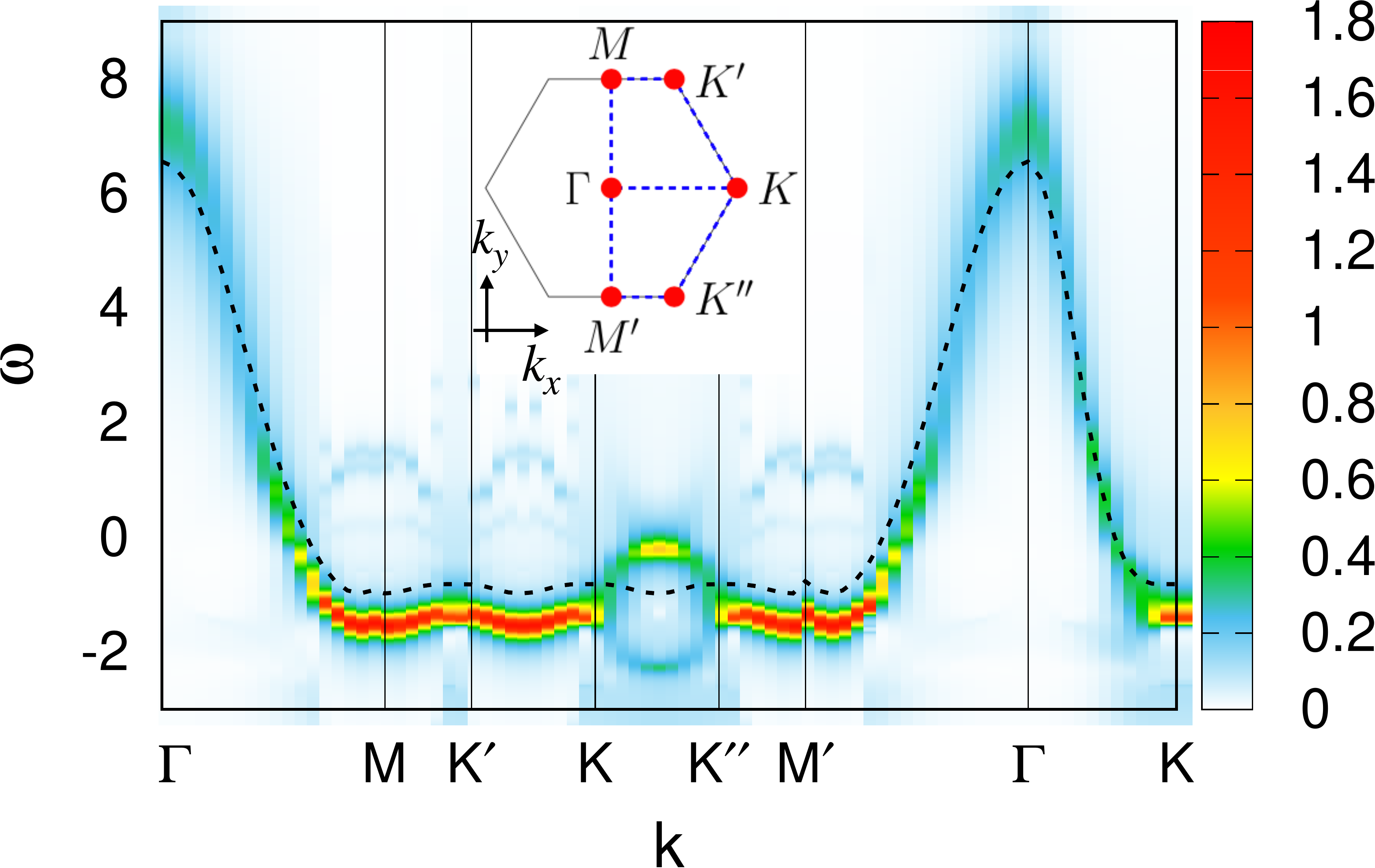}
    \caption{
    Momentum-dependent spectral function along high-symmetry momenta [see labels in the inset] in the PFB system with $t'=0.15$ for $\ave{n}=0.9$ and $U=4.5$. Dashed black lines in each panel represents the shifted noninteracting band structure~($\varepsilon_{\bolds{k}}-\mu$) with a chemical potential $\mu=-0.9$. (Inset) Hexagonal Brillouin zone for the triangular lattice with high symmetry points marked, i.e., $\Gamma$ at $(k_x,k_y) = (0,0)$, $K$ at $(4\pi/3,0)$, $K'$ at $(2\pi/3, 2 \pi \sqrt{3}/3)$, $K''$ at $(2\pi/3, -2 \pi \sqrt{3}/3)$, $M$ at $(0,2\pi \sqrt{3}/3)$, and $M'$ at $(0,-2\pi \sqrt{3}/3)$. $M$ and $M'$ are equivalent, in both of twofold (in the presence of the nematicity) and sixfold (in its absence) rotational symmetries. In the presence of sixfold symmetry, $K'$ and $K''$ are equivalent as well. 
    \label{fig:Awk}}
\end{figure}

\section{ Model }

The Hubbard Hamiltonian on the isotropic triangular lattice reads
\begin{equation}
    H = \sum_{\bolds{k}, \sigma} \varepsilon_{\bolds{k}} c^{\dagger}_{\bolds{k} \sigma}  c_{\bolds{k} \sigma} 
     +U \sum\limits_{i} n_{i \uparrow} n_{i \downarrow} - \mu \sum\limits_{i \sigma} n_{i \sigma},
     \label{eq:Ham}
\end{equation}
where $c^{\dagger}_{\bolds{k} \sigma}(c_{\bolds{k} \sigma})$ creates (annihilates) an electron with momentum $\bolds{k}=(k_{x}, k_{y})$ and spin $\sigma$ at site $i$, $n_{i \sigma} \equiv c^{\dagger}_{i \sigma} c_{i \sigma}$. The repulsive Hubbard interaction is denoted as $U (>0)$, and $\mu$ is the chemical potential. The non-interacting band dispersion for the triangular lattice is given as 
\begin{eqnarray}
   & \varepsilon_{\bolds{k}}(t,t') = -t \big[-2 \cos(k_{x}) -4 \cos(k_{x}/2) \cos(\sqrt{3} k_{y}/2) \big] \nonumber \\
    & +t' \big[-2  \cos(\sqrt{3} k_{y}) - 4\cos(3k_{x}/2) \cos(\sqrt{3} k_{y}/2) \big], \label{eq:disp}
 \end{eqnarray}
 where $t$ is the nearest-neighbor hopping (taken as a unit of energy) and $t'$ is the second-neighbor hopping. Here, we consider $(t,t')= (1.0,0.15)$, which
 possesses a nearly flat region along $K-K'-K''$, see dashed line in Fig.~\ref{fig:Awk}.
For the interaction, we set an intermediate 
$U=4.5t$, with the inverse temperature set to be $\beta\equiv1/(k_{B}T)=30/t$ except in Fig.~\ref{fig:xi_alpha}(c).

 \section{Numerical Method}
 
To study paramagnetic phases with no spin imbalance, we employ the dynamical mean-field theory~(DMFT) combined with the fluctuation exchange approximation~(FLEX), known as the FLEX+DMFT~\cite{Kitatani2015}. This method comprises DMFT and FLEX double loops, solved self-consistently at each FLEX+DMFT iteration. In this work, we solve the DMFT impurity problem by the modified iterative perturbation theory~\cite{Arsenault2012, Kajueter1996}. The momentum-dependent FLEX self-energy is constructed from the bubble and ladder diagrams. After removing the doubly-counted diagrams in the local FLEX self-energy, the FLEX+DMFT self-energy is updated. The momentum-dependent self-energy in the FLEX+DMFT incorporates vertex corrections generated from the DMFT iterations into the local part of the FLEX self-energy. Even though our FLEX+DMFT method does not deal with spatial vertex corrections, larger coordination number and frustrated magnetic fluctuations in the triangular lattice give rise to more local self-energies and less dominant spatial vertex corrections than in the square lattice~\cite{Vranic2020}. As a result, the FLEX+DMFT is considered to be a reliable approach that incorporates local and nonlocal correlations.

 When we start from the non-interacting tight-binding Hamiltonian,  Eq.~(\ref{eq:Ham}), that has the sixfold rotational~($C_{6}$) lattice symmetry, the solution of the many-body problem may exhibit a lower the symmetry. To study the phases with/without $C_{6}$ symmetry, we solve the FLEX+DMFT loops with/without imposing the $C_{6}$ constraints.
To explore the Pomeranchuk instability with the broken $C_{6}$ symmetry, we take an initial self-energy as $\Sigma_{\rm in} = 0.05 [\cos(k_{x}) − \cos(\sqrt{3}k_{y}/2) \cos(k_{x}/2) ]$ which acts as a seed for distorting the Fermi surface for the FLEX+DMFT iterations. 
The FLEX+DMFT calculations are here performed on a $64 \times 64$ momentum grid and an energy mesh with $2048$ points.

  \begin{figure}[t]
    \centering
    \includegraphics[width=0.9\columnwidth]{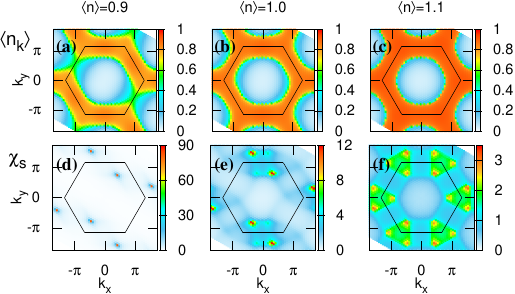}
    \caption{Momentum distribution functions~(top panels) and spin susceptibilities~(bottom) are displayed in momentum space for band fillings $\ave{n}$ = 0.9~(a, d), 1.0~(b, e), and 1.1~(c, f). All results are calculated for the PFB systems with $t' = 0.15$ and $U=4.5$. The black hexagon in each panel indicates the Brillouin zone.  Note different color bars between different band fillings.
    \label{fig:chis_nk_tp015}
    }
\end{figure}

\section{  Nematicity and non-Fermi liquid behavior}
We start with presenting the momentum  distribution function plotted in panels (a-c) in Fig.~\ref{fig:chis_nk_tp015}(top rows). For a system with a well-defined Fermi surface, $\ave{n_{k}}$ should take the value of unity~(zero) inside~(outside) the Fermi surface for $T\rightarrow0$. For all band fillings in our results, the maxima of the momentum-dependent distribution function are below unity.
The system exhibits a filling-dependent degrading of $C_{6}$ down to a twofold $C_{2}$ symmetry in $\ave{n_{k}}$. Namely, we have here an emergence of nematicity, or a {\it Pomeranchuk instability}. The breaking of 
 $C_{6}$ is seen to occur even right at half-filling, while the electron-doped case shows a preserved $C_{6}$.

To quantify the broken $C_{6}$ symmetry, we introduce point-group resolved Pomeranchuk order parameters defined as
$
    \xi_{d_{x^2-y^2}} =
    \sum_{\bolds{k}} 
   d_{x^2-y^2} (\bolds{k})
   n_{\bolds{k}}
$
 and
$
    \xi_{d_{xy}} =
    \sum_{\bolds{k}} 
    d_{xy}(\bolds{k}) n_{\bolds{k}},
$
with $\sum_{\bolds{k}}=1$~\cite{Kiesel2013}. The form factors, $d_{x^2-y^2}=\cos(k_x)-\cos(\sqrt{3}k_y/2)\cos(k_x/2)$ and $d_{xy}=\sqrt{3}\sin(\sqrt{3}k_y/2)\sin(k_x/2)$, describe the distortion of the Fermi surface in the point group $C_{6}$, and $\xi$ is a real number with values between zero (when $C_{6}$ is preserved) and unity. 

\begin{figure}[t]
    \centering
    \includegraphics[width=0.9\columnwidth]{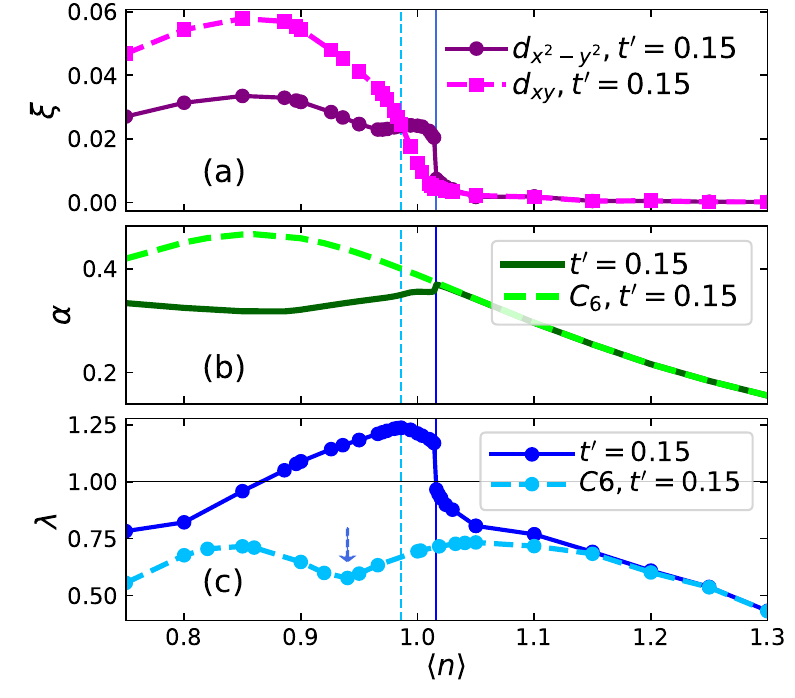}
    \caption{
   (a) Pomeranchuk order parameter $\xi$. $\xi_{d_{x^2-y^2}}$ for $t'=0.15$ is represented by purple lines. $\xi_{d_{xy}}$ is shown by magenta lines. (b) Exponent $\alpha$ of the impurity self-energy for systems with~(dashed curves) and without~(solid) imposed sixfold symmetry. 
   (c) The largest eigenvalue $\lambda$ of the linearized Eliashberg equation for the singlet pairing symmetry~ against band filling for the PFB system with $t'=0.15$ and $U=4.5$, with the broken~(dark blue curves) or unbroken~(sky-blue) sixfold symmetry.
    black horizontal line marks $\lambda=1$, and an arrow points to the dip in $\lambda$ when $C_{6}$ is enforced. 
    Vertical blue solid lines indicate $\ave{n_{\rm c_{1}}}$~(see text).
   Vertical dotted sky-blue lines are at $\ave{n_{\rm c_{1}}}=\ave{n_{\rm c_{2}}}$. 
    \label{fig:xi_alpha}}
\end{figure}

Figure \ref{fig:xi_alpha}(a) displays $\xi$ against the band filling.  
We can see that, as the band filling is reduced, $\xi$ starts to grow, and at a critical band filling $\ave{n_{{\rm c_{1}}}}=1.02$~[vertical blue line in Fig.~\ref{fig:xi_alpha}(a)], $\xi_{d_{xy}}$ undergoes a first-order phase transition~\cite{Yamase2005, Kitatani2017th}. 
At this filling, the onset of nematicity is accompanied by a Lifshitz transition, where the Fermi surface delineated by the ridges in $|G_{\bolds{k}}|^2$~(not shown) is not only distorted but undergoes a topological change from closed to open structures.

We further notice that the filling dependence of the nematicity differs between $\xi_{d_{x^{2}-y^{2}}}$ and $\xi_{d_{xy}}$ in the PFB model; compare purple and magenta lines in Fig.~\ref{fig:xi_alpha}(a). For $0.986<\ave{n} <\ave{n_{\rm c_{1}}} $, $\xi_{d_{x^{2}-y^{2}}}$ is dominant, while $\xi_{d_{xy}}$ takes over below $\ave{n}=0.986$, which we call the second characteristic band filling, $\ave{n_{{\rm c}_{2}}}$~[vertical dashed sky-blue line in Fig.~\ref{fig:xi_alpha}(a)].   
While $\xi_{d_{x^{2}-y^{2}}}$ displays a first-order transition at $\ave{n_{{\rm c_{1}}}}$, $\xi_{d_{xy}}$ exhibits a crossover at $\ave{n_{{\rm c}_{2}}}$. 
This suggests that thermodynamic parameters such as temperature at which  $\xi_{d_{xy}}$ and $\xi_{d_{x^{2}-y^{2}}}$ experience the first-order transitions are different from each other.

To trace back the origin of the nematic phases, let us next present the momentum-dependent spin susceptibility $\chi_{s}(\bolds{k})$ for the PFB model in Fig.\ref{fig:chis_nk_tp015}(d-f). In the electron-doped regime where the Pomeranchuk instability is absent, $\chi_{s}$ respects the sixfold rotational symmetry of the lattice, with peaks at $\bolds{k}=(\sqrt{3}\pi/2,0)$ and its equivalent positions under $C_6$. As band filling is decreased below the half-filling, the spin susceptibility develops spikes around $\bolds{k}=(\sqrt{3}\pi/3,2\pi/3)$ and the equivalent places under a $C_2$ subgroup of the original $C_{6}$ rotational symmetry. 
The appearance of spikes at mid-momenta in the spin susceptibilities indicates the presence of long-range spin fluctuations in our systems~\cite{Notespect}.

In general, an electronic nematicity without breaking the translational symmetry can be driven by structural transitions, charge~\cite{Massat2016} or spin~\cite{Fernandes2014} fluctuations. Our Hamiltonian does not deal with the distortion of the lattice or phonons and thus precludes structural transitions. We have checked that the charge susceptibility is at least an order of magnitude smaller than the spin susceptibility. Thus the spin-mediated correlations should be responsible for the emergence of the  Pomeranchuk instability~\cite{Kitatani2017}.

 \begin{figure}[t]
    \centering
    \includegraphics[width=0.9\columnwidth]{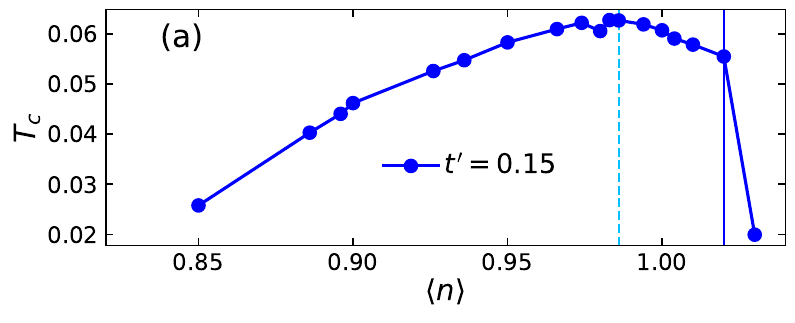}
    \includegraphics[width=0.9\columnwidth]{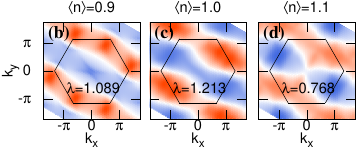}
    \caption{
    (a) Superconducting transition temperature $T_c$ against band filling for 
    the PFB.
    Vertical solid line indicates $\ave{n_{\rm c_{1}}}$, while vertical dotted line indicates $\ave{n}=\ave{n_{\rm c_{2}}}$. 
    (b-d) Gap functions in momentum-space with 
    singlet pairing for the PFB system with $t'=0.15$ for $U=4.5$ and $\beta=30$ for $\ave{n}=0.9$ (a), 1.0(b) and 1.1(c). Black hexagons indicate the Brillouin zone. Color code
for the gap function is bluish~(reddish) for negative~(positive)
values, for which we have omitted the color bars since the
linearized Eliashberg equation does not indicate magnitudes
of $\Delta$. 
    \label{fig:lambda_gapk}}
\end{figure}

Now let us turn to a non-Fermi liquid character of the present electronic systems, 
since the flat portions of the band may well exert peculiar effects.  We can quantify this in terms of the impurity self-energy in DMFT by fitting the imaginary part of the self-energy on Matsubara axis to $ |{\rm Im} \Sigma_{\rm DMFT} ({\rm i} \omega_{n})| \propto \omega_{n}^{\alpha}$, and present the result for the exponent  
$\alpha$ in Fig.~\ref{fig:xi_alpha}(b). In general, $\alpha= 1$ at small $\omega_{n}$
(c.f., $\alpha= 2$ on real frequency axis as $|{\rm Im} \Sigma_{\rm DMFT} ( \omega)| \approx \max(\omega^2, T^2)$ at small $T$) 
characterizes the Fermi liquid, while $\alpha < 0.5$ will signify a non-Fermi liquid~(bad metal) behavior~\cite{Werner2008, Vidhyadhiraja2009, Ishida2010, Werner2020}.
Above the first order Pomeranchuk transition for $\ave{n}>\ave{n_{{\rm c_{1}}}}$, $\alpha$'s computed for systems with~(dashed lines) and without~(solid lines) the enforced $C_{6}$ constraint trivially coincide with each other. We can see that both systems display strong non-Fermi liquid behavior with $\alpha$ well below $1$. If we turn to $\ave{n}<\ave{n_{{\rm c_{1}}}}$ for which we have revealed the nematicity, Fig.~\ref{fig:xi_alpha}(b) shows notable differences in $\alpha$ between the cases where $C_{6}$ is enforced or not. After a sharp drop at $\ave{n_{{\rm c_{1}}}}$ as the band filling is reduced, $\alpha$ gradually increases~(decreases) in the presence~(absence) of the imposed sixfold constraint. 
Eventually $\alpha$ starts to decrease with decreasing $\ave{n}$
at $\ave{n} \approx 0.85$ in the PFB system. The persistent $\alpha<0.5$ for $\ave{n}<\ave{n_{{\rm c_{1}}}}$ implies that the nematic phase resides in the non-Fermi liquid regime.

\section{Superconductivity}\label{subsec:sup}

 Now let us come to our key interest in pairing instabilities, for which we solve the linearized Eliashberg equation, 
$
    \lambda \Delta (k) = -\frac{1}{\beta} \sum_{k'}
    V_{\rm eff}(k-k') G_{k'} G_{-k'} 
    \Delta(k'),
$
to find the largest eigenvalue $\lambda$ for the spin-singlet, even-frequency superconducting gap function $\Delta$. Here, $k\equiv(\bolds{k}, {\rm i} \omega_{n})$ with $\omega_{n}$ the fermionic Matsubara frequency, 
and the effective interaction for singlets given as $ V_{\rm eff}(k) = U+3 U^2 \chi_{s} (k) /2 - U^2 \chi_{c} (k)/2$. The pairing is identified when $\lambda$ exceeds unity~\cite{eigp}. 
Figure~\ref{fig:xi_alpha}(c) depicts $\lambda$ in the presence~(dashed green lines) or absence~(solid blue) of imposed $C_{6}$ symmetry in PFB.

 When the sixfold rotational symmetry is enforced, we get $\lambda<0.8$ in PFB model, indicating that the singlet superconductivity does not arise 
 for the temperature ($k_BT = t/30$) considered here. We can still notice that $\lambda$ displays a double-peak structure with a minimum at $\ave{n}_{\rm min}=0.95$. The dip is shown to occur at the band filling at which the $d_{x^2-y^2}$ gap function with two-nodal lines for $\ave{n}>\ave{n}_{\rm min}$ changes into a more complicated multi-nodal-line gap functions for $\ave{n}<\ave{n}_{\rm min}$, see ~\cite{Extendpair,Suppmat} for details. This behavior of the gap function reflects a crossover from the antiferromagnetic spin structure with a single nesting vector for $\ave{n}>\ave{n}_{\rm min}$, to a more complex spin structure for $\ave{n}<\ave{n}_{\rm min}$ where single peaks in the spin susceptibility evolve into extended structures [see  Fig.\ref{fig:chis_nk_tp015}(d-f)]. Thus the system for $\ave{n}<\ave{n}_{\rm min}$ goes {\it beyond} the conventional nesting physics. Similar structure in $\lambda$ and associated gap function have also been reported for PFB systems on the square lattice~\cite{Sayyad2020}, again in the absence of nematicity.

In a dramatic contrast, if we allow the $C_{6}$ symmetry to be broken 
spontaneously, 
$\lambda$ soars from those with $C_{6}$ restriction, as seen  
for $\ave{n_{\rm c}} < \ave{n} <1.15$. This occurs concomitantly with the Pomeranchuk order parameters ($\xi$'s), which grow precisely in this filling region. Just below $\ave{n_{\rm c}}$, $\lambda$ in the systems with broken $C_{6}$ (solid blue lines in Fig.~\ref{fig:xi_alpha}~(c)) exhibits a rapid growth and exceeds unity. This is inherited in the superconducting transition temperatures~$(T_{\rm c})$, presented in Fig.~\ref{fig:lambda_gapk}(a).

 $T_{\rm c}$, with the broken $C_6$, exhibits a single-dome structure as a function of band filling. We can observe that the presence of a flat portion in PFB or a van Hove singularity for $t'=0$ have similar effects on the largest values of $T_{\rm c}$ when $\xi_{d_{xy} } \geq \xi_{d_{x^2-y^2}}$. One should note that, while a van Hove singularity at $E_F$ only occurs at a single point on the filling axis, a flat portion of the band can accommodate a range of band filling. This difference is reflected in the width of the $T_{\rm c}$ dome at a given temperature; see Fig.~\ref{fig:lambda_gapk}(a) and SM~\cite{Suppmat}. The maximum of $T_{\rm c}$ in the PFB is seen to take place close to $\ave{n_{\rm c_{2}}}$ 
 at which $\xi_{d_{xy} } $ exceeds $ \xi_{d_{x^2-y^2}}$. 
 Note that the superconducting transition temperature becomes almost doubled as we pass through $\ave{n_{\rm c_{1}}}$, see Fig.~\ref{fig:lambda_gapk}(a), which should come from the interplay between nematicity, spin fluctuations, and superconductivity. 

Let us now delve into the gap function in momentum space in Fig.~\ref{fig:lambda_gapk}(b-d). In the electron-doped regime, the PFB model exhibits a conventional $d_{x^2-y^2}$ paring symmetry~\cite{Gapsym}. This behavior of the gap function persists for $\ave{n} >\ave{n_{\rm c}}$.
On the other hand, below $\ave{n_{\rm c_{1}}}$ where the $C_{6}$ symmetry is broken down to its $C_{2}$ subgroup, the dominant channel of instability is a mixture of  $s_{x^2+y^2}$, $d_{x^2-y^2}$ and $d_{xy}$-wave symmetries, see Fig.S16 
in SM~\cite{Suppmat}.

To better understand the role of nematicity in superconducting phases, we can look at $\Delta V_{\rm eff} = V_{\rm eff}- V_{\rm eff}^{C_{6}}$, where $V_{\rm eff}^{C_{6}}$ is the effective interaction with the imposed $C_{6}$ constraint.  As shown in Fig.~S19 in SM~\cite{Suppmat}, $V_{\rm eff}$ is much {\it intensified} when $C_6$ is lifted. 
Since $\chi_{c}$ is much smaller than $\chi_{s}$, the effective pairing interaction reflects the momentum-dependence of the spin-susceptibility under the Pomeranchuk distortions.
This effective interaction assists electrons to nonlocally form Cooper pairs~\cite{Halboth2000,Yamase2010,Kitatani2017th}.
 The deformation in $\Delta V_{\rm eff}$ allows first-order perturbation corrections in the distortion, which should be responsible for the drastic changes in $\lambda$ below $\ave{n_{\rm c}}$. This contrasts with the previous study on the interplay between nematicity and superconductivity, where the enhancement of $\lambda$ originates from the second-order perturbation corrections and thus results in much smaller changes~\cite{Kitatani2017}.

\section{ Discussion and summary}\label{sec:conclution}

We have studied whether and how an emergent nematicity affects superconductivity in partially flat bands on the regular triangular lattice. We have shown with the FLEX+DMFT that nematicity dramatically affects pairing symmetry, and the $T_C$ is significantly enhanced by the lowered point-group symmetry in the electronic structure. This is shown to occur in a non-Fermi liquid regime, which is characterized by blurred Fermi surfaces, momentum-dependent fractional occupations of the band, and a fractional power-law in the self-energy. In the presence of nematic order, the superconducting symmetry changes from an (extended) $d_{x^2-y^2}$-wave to a $s_{x^2+y^2} -d_{x^2-y^2}-d_{xy}$-wave, where unlike the conventional nesting-driven case, the pairing interaction is governed by an intricate spin susceptibility structure.

Future works should include the elaboration of the way in which the non-Fermi liquid property affects the superconductivity, and exploration of the interplay between Pomeranchuk instability and superconductivity in multi-band/orbital systems with flat regions.

\section{ Acknowledgements}

Sh.S. acknowledges financial support from the ANR under the grant ANR-18-CE30-0001-01~(TOPODRIVE) and the European Union Horizon 2020 research and innovation program under grant agreement No.829044~(SCHINES). M.K was supported by JSPS KAKENHI Grand Numbers JP20K22342 and JP21K13887. H.A. thanks CREST (Core Research for Evolutional Science and Technology) 
``Topology” project (Grant Number JPMJCR18T4) from Japan Science and Technology Agency, and JSPS KAKENHI Grant JP17H06138.

\newpage
\appendix
\onecolumngrid

\setcounter{secnumdepth}{5}
\renewcommand{\theparagraph}{\bf \thesubsubsection.\arabic{paragraph}}

\renewcommand{\thefigure}{S\arabic{figure}}
\setcounter{figure}{0}

\section{Observables}\label{sec:observ}
Let us define the observables.

\textbf{DMFT and momentum-dependent spectral functions:}
The DMFT spectral function, $A(\omega)=-1/\pi \Im[G_{\rm imp}( \omega)]$, is computed by analytically continuing the DMFT impurity Green's function on 
Matsubara frequency axis with the Pad\'e approximation~\cite{Schott2016}. Similarly, the momentum-dependent FLEX spectral function, $A( \bolds{k}, \omega)=-1/\pi \Im[G({\bolds k}, \omega)]$, is evaluated from the momentum-dependent Green's function with an analytical continuation.

\textbf{Normalized double-occupancy,}
$\ave{n_{\uparrow} n_{\downarrow}}/(\ave{n_{\uparrow} } \ave{ n_{\downarrow}}) $), is the ratio between the number of doubly occupied sites~$\ave{n_{\uparrow} n_{\downarrow}}$ to the uncorrelated value $\ave{n_{\uparrow}} \ave{n_{\downarrow}}$.

\textbf{Momentum-dependent distribution function}
is given by
\begin{equation}
   \ave{ n_{\bolds{k}} } = 
   \frac{1}{2} 
    \sum\limits_{\sigma} 
    \ave{c^{\dagger}_{\bolds{k}\sigma} c_{\bolds{k} \sigma}}.
\end{equation}

\textbf{Pomeranchuk order parameters,}
$\xi_{d_{x^2-y^2}}$ and $\xi_{d_{xy}}$, which quantify the amount of symmetry breaking~\cite{Yamase2005}, are given on the triangular lattice as~\cite{Kiesel2013}
\begin{align}
    \xi_{d_{x^2-y^2}} &=
    \sum\limits_{\bolds{k}} 
 d_{x^2-y^2} (\bolds{k})
 n_{\bolds{k}},
\\
    \xi_{d_{xy}} &=
    \sum\limits_{\bolds{k}} 
    d_{xy}(\bolds{k})
 n_{\bolds{k}},
\end{align}
where $\sum\limits_{\bolds{k}}=1$.
The form factors, $d_{x^2-y^2}=
\cos(k_x)- \cos(\sqrt{3}k_y/2)\cos(k_x/2)$ and $d_{xy}=\sqrt{3}\sin(\sqrt{3}k_y/2)\sin(k_x/2)$, describe the 
component-resolved distortion of $d$-wave instabilities for the Fermi surface distortion in the point group $C_{6}$.
When the $C_{6}$ symmetry is preserved, both $  \xi_{d_{x^2-y^2}} $ and $ \xi_{d_{xy}}$ are equal to zero.

\textbf{Static spin and charge susceptibilities.} 
The static spin susceptibility reads
\begin{equation}
    \chi_{s}(\bolds{k}) = 2 \int_{0}^{\beta} {\rm d} \tau 
    \ave{S^{z}_{\bolds{k}}(\tau) S^{z}_{-\bolds{k}} (0)},
\end{equation}
where $\tau$ denotes imaginary time and $\beta$ is the inverse temperature, while the static charge susceptibility is given by
\begin{equation}
    \chi_{c} (\bolds{k}) = \int_{0}^{\beta} {\rm d} \tau 
    \ave{n_{\bolds{k}}(\tau) n_{-\bolds{k}} (0)}.
\end{equation}
The local spin susceptibility is then given by
\begin{align}
    \chi_{s}^{\rm loc} =\sum_{\bolds{k}} \chi_{s}(\bolds{k}),
\end{align}
while the DMFT impurity spin susceptibility is calculated as
\begin{align}
\chi^{\rm imp}_{s} = \frac{\chi_{0}(0)}{ 1- U \chi_{0}(0)},
\end{align}
where the polarization function is $\chi_{0}(0)= - \sum_{\omega_{n}} G^{\rm imp} ({\rm i} \omega_{n})  G^{\rm imp} (-{\rm i} \omega_{n})$, with $G^{\rm imp}$ being the  DMFT impurity Green's function and $\sum_{\omega_{n}}=1$.

\textbf{Exponent of the impurity self-energy.}
To quantify non-Fermi liquid behavior we can fit the imaginary parts of the DMFT self-energies at small Matsubara frequencies~($\omega_{n}$) to
\begin{align}
    |{\rm Im} \Sigma_{\rm DMFT} ({\rm i} \omega_{n})| &\propto \omega_{n}^{\alpha}.
\end{align}
For Fermi liquids, ${\rm Im} \Sigma({\rm i} \omega_{n}) \propto {\rm i} \omega_{n}$ 
with the exponent $\alpha=1$ on the Matsubara axis (i.e., ${\rm Im} \Sigma(\omega) \propto \omega^2$ 
on the real frequency axis). If $\alpha< 0.5$, the bad metallic behavior is indicated as a signature of non-Fermi liquid.

\textbf{Superconducting gap functions.}
For pairing instabilities we solve the linearized Eliashberg equation to find leading eigenvalues $\lambda$ and superconducting gap functions $\Delta$ as
\begin{equation}
    \lambda \Delta (k) = -\frac{1}{\beta} \sum\limits_{k'}
    V_{\rm eff}(k-k') G(k') G(-k' ) 
    \Delta(k'),
\end{equation}
where $k=(\bolds{k}, {\rm i} \omega_{n})$ with the fermionic Matsubara frequency $\omega_{n}$, $G$ denotes the single-particle Green's function, 
and the effective interaction for the spin-singlet
pairing, 
\begin{align}
    V_{\rm eff}(k) &= U+\frac{3}{2} U^2 \chi_{s} (k) -\frac{1}{2} U^2 \chi_{c} (k)
.
\end{align} 
In this paper, we focus ouselves on even-frequency gap functions with even parity in the Matsubara frequency,  which satisfy $\Delta(\bolds{k}, {\rm i} \omega_{n}) =\left[ \Delta(\bolds{k}, {\rm i} \omega_{n}) \right. +\left. \Delta(\bolds{k}, -{\rm i} \omega_{n}) \right]/2$. Superconductivity is identified when the largest eigenvalue $\lambda$ exceeds unity.

\section{Ideal partially flat-band model}~\label{app:ipfb}

\begin{figure}[t]
    \centering
    \includegraphics[width=0.58\columnwidth]{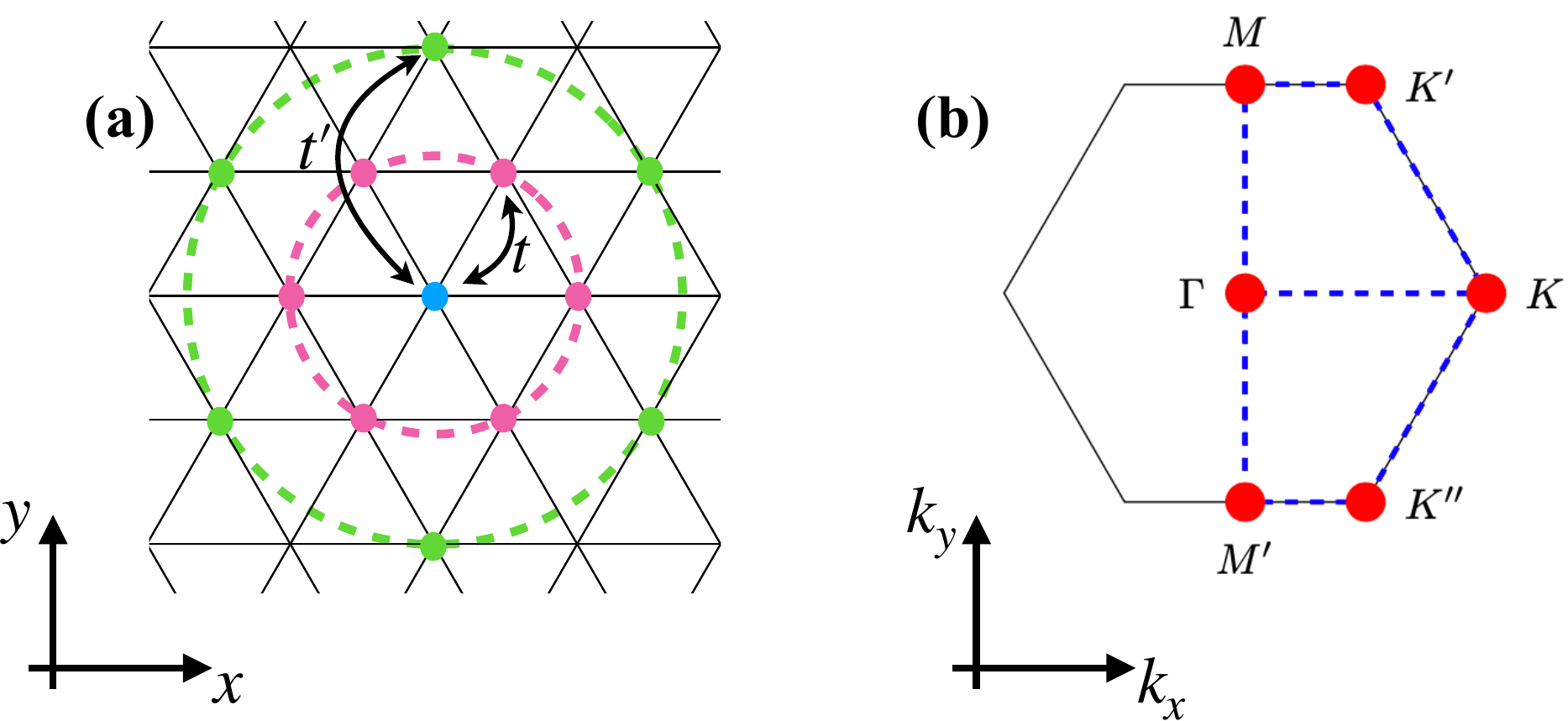}
    \caption{
    (a) Triangular lattice. Hopping from 
    a lattice point (blue dot) to the nearest neighbors~(pink) with an amplitude $t$, 
    and to the second neighbors~(green) with $t'$. 
    (b) Hexagonal Brillouin zone for the triangular lattice with high symmetry points marked. $M$ and $M'$ are equivalent, in twofold (in the presence of the nematicity) as well as sixfold (in its absence) rotational symmetries. In the presence of sixfold symmetry, $K'$ and $K''$ are equivalent.
   \label{fig:lattice}}
\end{figure}

In this section, we explore how the way in which the flat portion in the band dispersion is prepared affects the physics.  So here we introduce the Hubbard model on a model [which we call the ideal partially flat-band~(iPFB)] with a truncated dispersion given as
\begin{align}
    \varepsilon_{\bolds{k}}^{\rm iPFB} = -\frac{W}{W_{\kappa}} \times
    \begin{cases}
    \varepsilon_{\bolds{k},0} & \quad \varepsilon_{\bolds{k},0} <\kappa ,\\
    \kappa & \quad \text{otherwise}.
    \end{cases}
\end{align}
Namely, the dispersion is made perfectly flat for 
energies below $\kappa$ for a tight-binding model 
having a nearest-neighbor $t$ and second-neighbor $t'$ 
with the dispersion,
\begin{align}
    \varepsilon_{\bolds{k},0} =& \;-t \big[2 \cos(k_{x}) +4 \cos(k_{x}/2) \cos(\sqrt{3}k_{y}/2 ) \big] \nonumber \\
    & - t' \big[2  \cos(\sqrt{3} k_{y}) + 4 \cos(3k_{x}/2) 
     \cos(\sqrt{3}k_{y}/2) \big], \label{eq:disp}
 \end{align}
see Fig.~\ref{fig:lattice}(a). 
Thus the larger the $\kappa$, the more extended the flat region.   
We have here inverted the signs of $t$ and $t'$ 
to put the van Hove singularity near the band 
bottom for the triangular lattice ~(occupied at 
small band fillings).   The band width is that of 
Eq.(C2) 
minus $\kappa$.  Here we take the band 
width to be $W=7.533t$.

\subsection{Results for iPFB}
 
Let us present the numerical result for iPFB, 
here for $U=4.5 t$ and, except in  Figs.~\ref{fig:docc_ipfb}(d), \ref{fig:lambda_ipfb}(b), \ref{fig:xi_chiimp}(a), the inverse temperature is $\beta=1/(k_{B}T)=30/t$. We take $t$ as the unit of energy as in the main text.
    
\subsubsection{Nematicity and non-Fermi liquid behavior}

\begin{figure}[t]
    \centering
    \includegraphics[width=0.75\columnwidth]{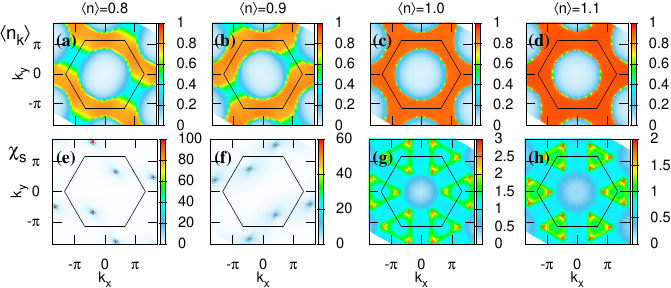}
    \caption{
    For the iPFB model, here with $\kappa = 1.0$ and $U=4.5$, momentum distribution functions~(top panels) and spin susceptibilities~(bottom) are presented in momentum space for band fillings $\ave{n}$ = 0.8~(a, e, i), 0.9~(b, f, j), 1.0~(c, g, k) and 1.1~(d, h, l). The black hexagon in each panel represents the Brillouin zone. Note different color codes for different band fillings.
    \label{fig:chis_nk_eps1}}
\end{figure}

\begin{figure}[t]
    \centering
    \includegraphics[width=0.75\columnwidth]{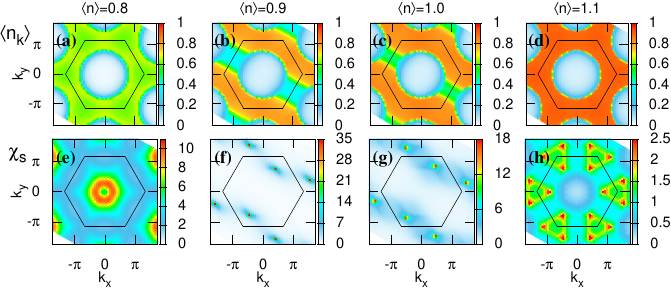}
    \caption{
    The same as the previous figure, for an increased $\kappa = 1.5$.
    \label{fig:chis_nk_eps15}}
\end{figure}

In search for footprints of non-Fermi liquid behavior, we start with the absolute value of Green's functions $|G_{\bolds{k}}|^{2}$~(top rows) and the momentum-dependent occupation number~(middle) in Figs.~\ref{fig:chis_nk_eps1}, \ref{fig:chis_nk_eps15} for the iPFB models with $\kappa=$ $1.0$ and $1.5$.  
The iPFB model displays an absence of sharp peaks in $|G_{\bolds{k}}|^{2}$ as for $t'=0$ (not shown), 
i.e., absence of well-defined Fermi surfaces, which hints that non-Fermi physics governs these systems. A further signature of non-Fermi liquid characters is evident in $\ave{n_{\bolds{k}}}$ in which we observe ${\rm max}[ \ave{n_{\bolds{k}}}]<0.95$ for all presented band fillings in the iPFB model.

In the hole-doped regime of the iPFB model with $\kappa=1.0$, we observe a broken sixfold symmetry, see panels (a,b) and (e,f) in Fig.~\ref{fig:chis_nk_eps1}. For $\kappa=1.5$ in Fig.~\ref{fig:chis_nk_eps15}, 
the reduction from $C_{6}$ to $C_{2}$ occurs even at half-filling, while the symmetry is resumed for smaller band fillings, namely $\ave{n}<0.8$. These results suggest that the onset of broken $C_{6}$ symmetry is shifted to higher band fillings as we increase the size of the flat portion. Thus we can conclude that the spontaneous 
nematicity occurs both in 
the PFB and iPFB, even though the 
detail of band dispersion is significantly 
different between the two models.

\begin{figure}[t]
    \centering
    \includegraphics[width=\columnwidth]{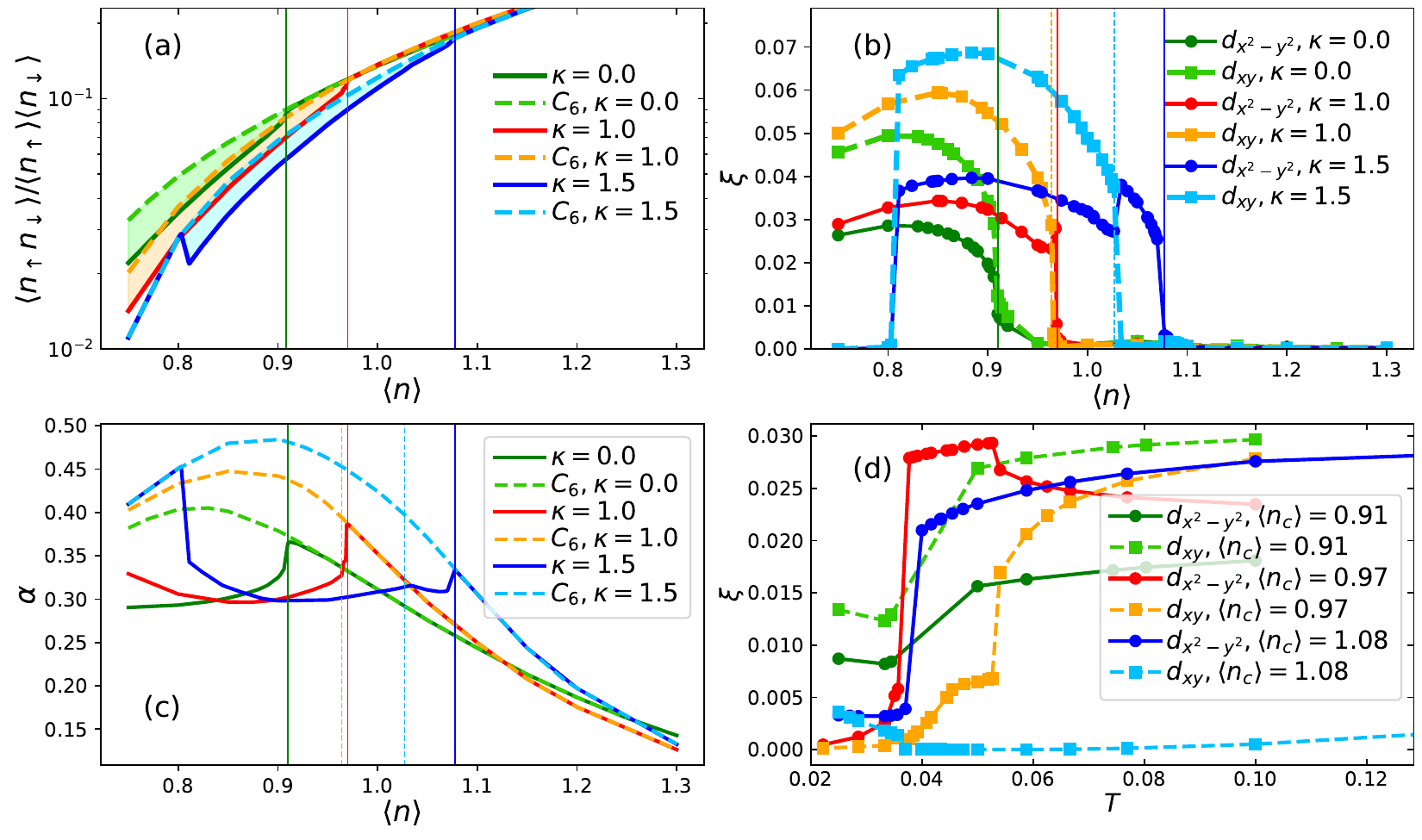}
    \caption{
    For iPFB (a) normalized double occupancy against the 
    band filling for various 
    values of $\kappa$  with~(dashed curves) and without~(solid curves) enforced $C_{6}$ symmetry. 
    Shaded areas mark the deviation between 
    the two cases. 
    (b) Pomeranchuk order parameters $\xi$ for various 
    values of $\kappa$.  (c) The exponent $\alpha$ 
    of the impurity self-energy at $\beta=30$. 
    Dashed curves represent the case of enforced $C_{6}$. 
    We take $\beta=30$ for these panels.  
    (d) The Pomeranchuk order parameters at respective  $\ave{n}=\ave{n_{\rm c}}$ against temperature 
    for various values of $\kappa$ (with the same color code as in (a)).  
    Vertical solid~(dashed) lines mark $\ave{n_{\rm c}}$~($\ave{n_{\rm c_{2}}}$), respectively.
    \label{fig:docc_ipfb}
    }
\end{figure}

If we look at the normalized double occupancy against band filling in Fig.~\ref{fig:docc_ipfb}(a) 
to see how this quantity is correlated with the spontaneous breaking of $C_{6}$.
Above a particular filling, the normalized double occupancy in systems with~(dashed curves) and without~(solid curves) enforced $C_{6}$ symmetry are identical. Below this critical band filling, which we call $\ave{n_{\rm c}}$, the Pomeranchuk instability occurs as a first-order phase transition. Figure~\ref{fig:docc_ipfb}(a) shows that $\ave{n_{\rm c}}$ (marked as vertical lines for the respective 
values of $\kappa$) depends systematically on the size of the flat portion as $\ave{n_{\rm c}}= 0.91$ for $\kappa=0$, $\ave{n_{\rm c}}=0.97$ for $\kappa=1.0$, and $\ave{n_{\rm c}}=1.08$ for $\kappa=1.5$.  
For $ \kappa=1.5$, if we further decrease the filling 
to $\ave{n}<0.8$ the $C_{6}$ symmetry is restored as we have already noted in Fig.S3. We refer to $\ave{n_{\rm cl}}=0.81$ as the lower critical filling.

We can also note that we have a Lifshitz transition at the critical band filling $\ave{n_{\rm c}}$, i.e., the closed ridges in $|G_{\bolds{k}}|^2$ above $\ave{n_{\rm c}}$ are topologically changed to an open structure, c.f. panels(b,c) in Fig.~\ref{fig:chis_nk_eps1} and (c,d) in Fig.~\ref{fig:chis_nk_eps15}. The second Lifshitz transition occurs at the lower $\ave{n_{\rm cl}}$, below which the open ridges in $|G_{\bolds{k}}|^2$ resume a closed structure, see Fig.~\ref{fig:chis_nk_eps15}(a). 

 \begin{figure}[t]
    \centering
    \includegraphics[width=\columnwidth]{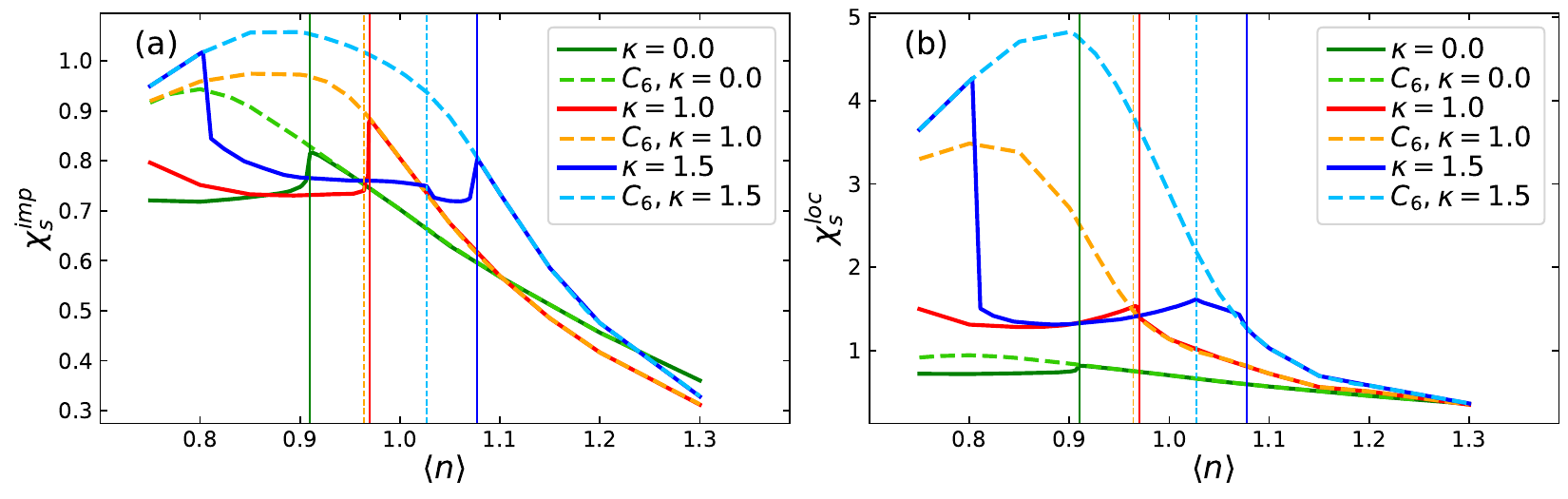}
    \caption{
    Impurity spin susceptibility~(a) and local spin susceptibility~(b) for the regular band model with $\kappa=0$~(green lines), and for the iPFB models with $\kappa=$ $1.0$~(red lines) or $1.5$~(blue lines). Dashed lines represent the results with the imposed sixfold symmetry. Vertical solid~(dashed) lines mark $\ave{n_{\rm c}}$~($\ave{n_{\rm c_{2}}}$).
    \label{fig:lchi_ipfb}}
\end{figure}

If we actually examine in Fig.~\ref{fig:docc_ipfb}(b) the Pomeranchuk order parameters~($\xi$s) against band filling for $\kappa$ increased from $0$ to $1.5$ at $\beta=30$. For $\kappa=0$, $\xi_{d_{xy}}$~(light green squares) and $\xi_{d_{x^2-y^2}}$~(dark green circles) experience a first-order phase transition at a $\ave{n_{\rm c}}$. Below this critical filling, both orders grow, with $\xi_{d_{xy}}$ larger of the two.
For $\kappa=1.0$, we observe a sharp rise in $\xi_{d_{x^2-y^2}}$ at a shifted $\ave{n_{\rm c}}$ accompanied by a sharp rise in $\xi_{d_{xy}}$ at $\ave{n_{\rm c2}}=0.964$~(vertical dotted orange line).  
When $\kappa$ is increased to $1.5$, the critical band filling $\ave{n_{\rm c}}$ for $\xi_{d_{x^2-y^2}}$ further increases, while $\xi_{d_{xy}}$ has a jump at $\ave{n_{\rm c_{2} }}=1.027$~(vertical dotted sky-blue line) accompanied by a drop in $\xi_{d_{x^2-y^2}}$. For $\ave{n}<\ave{n_{\rm c_{2}}}$, the dominant Pomeranchuk instability has $d_{xy}$ character. This continues until the lower critical filling $\ave{n_{\rm cl}}=0.81$ below which the iPFB system retrieves the $C_{6}$ symmetry and all instabilities are gone.

We now look at their temperature dependence of $\xi$s at respective $\ave{n_{\rm c}}$ in Fig.~\ref{fig:docc_ipfb}(d). For $\kappa=0$, the rapid growth of both $\xi_{d_{x^2-y^2}}$ and $\xi_{d_{xy}}$ simultaneously occur around $T\approx 0.05$. It is evident that these nematic orders at $\ave{n_{\rm c}}$ survive at higher temperatures with $\xi_{d_{xy}}$ remaining  dominant. For $\kappa=1.0$, we see a first-order phase transition in $\xi_{d_{x^2-y^2}}$ at $T\approx 0.038$ followed by a sharp increase in $\xi_{d_{xy}}$ at $T\approx 0.055$. At $T=0.066$, the two Pomeranchuk instabilities become comparable.  
For $\kappa=1.5$, we witness a first-order phase transition for $\xi_{d_{x^2 -y^2}}$ at $T\approx 0.038$, dark blue line in Fig.~\ref{fig:docc_ipfb}(d). This order parameter remains the only instability at higher temperatures as at $\ave{n_{\rm c}}$, with $\xi_{d_{xy}}$ almost vanishing as the temperature is increased.

We turn to the spin susceptibility in Figs.~\ref{fig:chis_nk_eps1}, \ref{fig:chis_nk_eps15}(bottom rows)  for $\kappa=$ $1.0$ and $1.5$, respectively. In the electron-doped regime where the Pomeranchuk distortion is absent, the spin susceptibility exhibits spikes corresponding to  antiferromagnetic spin structure at $\bolds{k}=(\sqrt{3}\pi/2,0)$  with it's equivalent sixfold k-points. Below $\ave{n_{\rm c}}$, the spin susceptibility exhibits two streaks, reflecting the $C_{2}$ symmetry. These are centered around 
$\bolds{k}= (\sqrt{3}\pi/3,2\pi/3)$, and its twofold equivalent places. At $\ave{n}=0.9$ for $\kappa=1.0$, the streaks in the spin susceptibility are rotated by $\pi/6$ and located at $\bolds{k}= (-\sqrt{3}\pi/3,2\pi/3)$, and its twofold equivalent places.  For $\kappa=1.5$, 
we can confirm that the sixfold symmetry is resumed in the spin susceptibility as well below $\ave{n}=\ave{n_{\rm cl}}$.  In agreement with the conclusion of the main text, we can thus infer that the nematicity is triggered by spin fluctuations.

 We can further look at the impurity and local spin susceptibilities in Fig.~\ref{fig:lchi_ipfb}(a,b), respectively. Right at $\ave{n}=\ave{n_{\rm c}}$, the $\chi_{s}^{\rm imp}$ and $\chi_{s}^{\rm loc}$ in systems without the enforced $C_{6}$ symmetry exhibit drastic deviations from the $C_{6}$-imposed results. At $\ave{n_{\rm c_{2}}}$ with $\kappa=$ $1.0$ and $1.5$ where the dominant instability changes from $\xi_{d_{x^2-y^2}}$ to $\xi_{d_{xy}}$, the impurity and local spin susceptibilities exhibit small kinks.

 In addition, at fillings below $\ave{n_{\rm c}}$~(solid vertical lines in Fig.~\ref{fig:lchi_ipfb}), the difference between $\chi^{\rm loc}_{s}$ in the presence~(dashed lines) and absence~(solid) of $C_{6}$ becomes more pronounced as the flat-band region is widened. This is as expected, since the many-body effects, reflected in $\chi_{s}(\bolds{k})$, are stronger in partially flat band systems. This property has also been noticed for partially flat-band systems studied with the determinant Monte-Carlo~\cite{Huang2019} and  FLEX+DMFT~\cite{Sayyad2020} calculations on the square lattice. 

  \begin{figure}[t]
    \centering
      \begin{tabular}{@{}p{0.32\linewidth}@{\,}p{0.32\linewidth}@{}@{\,}p{0.32\linewidth}@{}}
     \includegraphics[width=\linewidth]{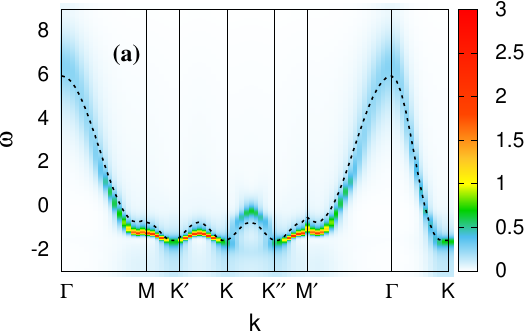}&
     \includegraphics[width=\linewidth]{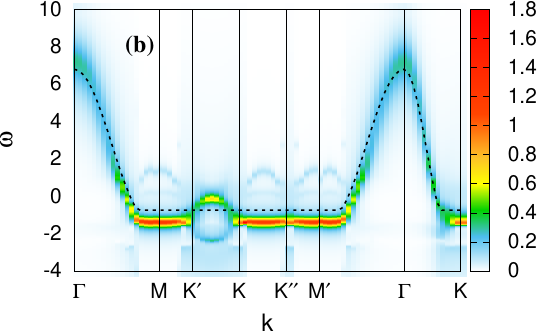}&
     \includegraphics[width=\linewidth]{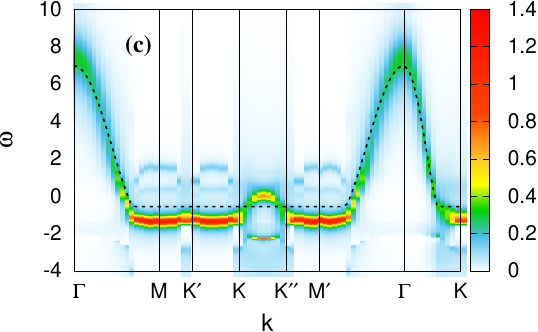}
  \end{tabular}
    \caption{
 Momentum-dependent spectral function along high-symmetry momenta for $\kappa=1.0$~(b) or $1.5$~(c) at $\ave{n}=0.9$. 
Dashed black line in each panel represents the noninteracting band structure shifted by the chemical potential $\mu=-0.73$~(b), or $-0.91$~(c).
    \label{fig:Awk_ipfb}}
\end{figure}

To further track the fingerprints of the nematic orders, we present in Fig.~\ref{fig:Awk_ipfb} the momentum-dependent spectral function, $A(\bolds{k}, \omega)$, in iPFB with $\kappa=1.0$~(b) or $\kappa=1.5$~(c) at $\ave{n}=0.9$. 
In the presence of the interaction, the flat region at the bottom of the bands seen in $A(\bolds{k})$ changes little, aside from a downward energy shift, except along the high symmetry lines where $C_{6}$ is broken as identified in Figs.~\ref{fig:chis_nk_eps1}, \ref{fig:chis_nk_eps15}. More specifically, for $\kappa=1.0$, the nematicity vector is along $K' \rightarrow K$, which changes to $K \rightarrow K''$ for $\kappa=1.5$, see labels in Fig.~\ref{fig:lattice}(b).  For the momenta in these regions, the spectral function splits with a gap $\lesssim U/2$.

 \begin{figure}[t]
    \centering
    \includegraphics[width=\columnwidth]{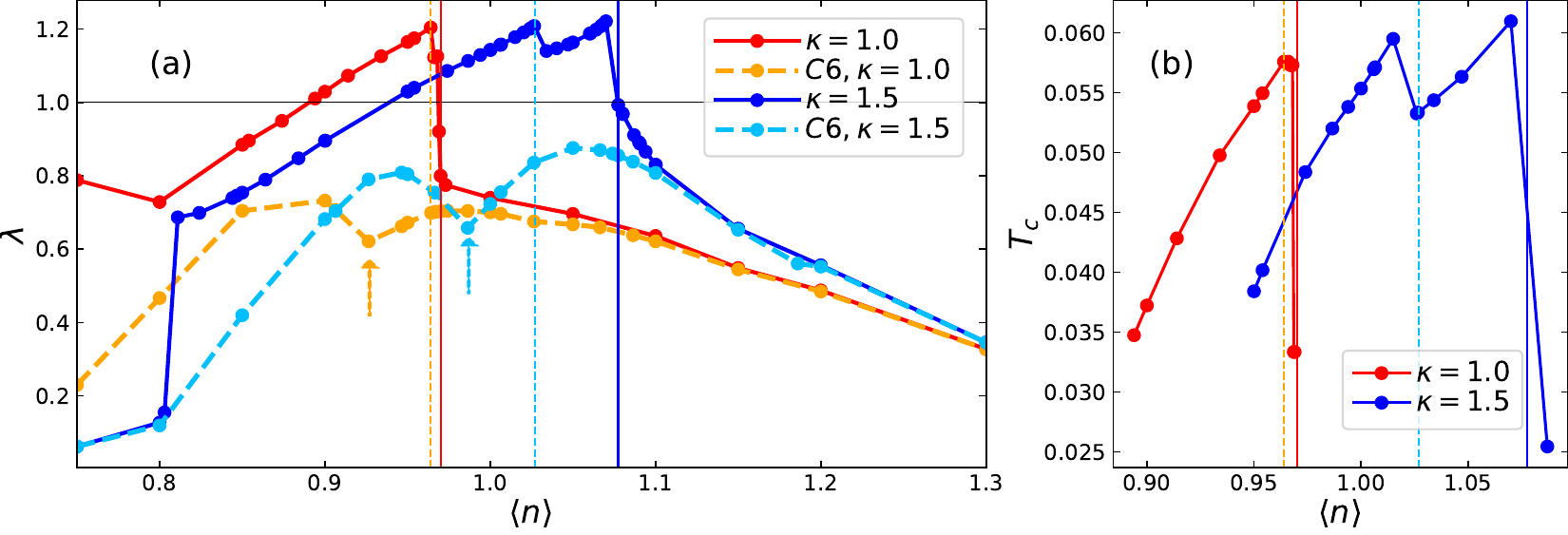}
    \caption{ 
    For the iPFB with $\kappa=1.0$~(red lines) or $\kappa=1.5$~(dark blue) the linearized Eliashberg eigenvalue $\lambda$~(a) and superconducting transition temperature $T_c$~(b) are plotted. In (a) dashed lines show $\lambda$ when the sixfold symmetry is imposed, the black horizontal line marks  $\lambda=1$, and arrows point to respective dips when the $C_{6}$ constraint is enforced.  Vertical solid lines indicate respective $\ave{n_{\rm c}}$, while dotted vertical lines mark respective $\ave{n}=\ave{n_{\rm c_{2}}}$.
    \label{fig:lambda_ipfb}}
\end{figure}

To corroborate the non-Fermi liquid character of electrons in our models, we plot the exponent $\alpha$ of the impurity self-energies in systems with~(dashed lines) and without~(solid) the sixfold symmetry in Fig.~\ref{fig:docc_ipfb}(c) above. For the whole 
band-filling
region studied, $\alpha$ exhibits values less than $0.5$, indicating non-Fermi liquids. In the electron-doped regime for $\ave{n}>\ave{n_{\rm c}}$ with unbroken $C_{6}$, $\alpha$ grows as we approach $\ave{n_{\rm c}}$ as the band filling is decreased. Just below $\ave{n_{\rm c}}$, $\alpha$ exhibits a sharp drop, followed by a gradual enhancement accompanying the Pomeranchuk instabilities. Close to $\ave{n}=\ave{n_{\rm c_{2}}}$, $\alpha$ in the presence of nematicity~(solid lines) undergoes another drastic reduction.

\subsubsection{Superconductivity}

We now turn to superconductivity in the iPFB.  We present the largest eigenvalue $\lambda$ of the Eliashberg equation for the singlet pairing in Fig.~\ref{fig:lambda_ipfb}(a). If we first look at the result when the $C_{6}$ symmetry is imposed~(dashed lines), $\lambda$ remains below $0.85$, i.e., no superconducting instabilities. Similar to our finding in the main text for the PFB model, we can characterize the double-dome structure in $\lambda$ as a signature of a change in the number of nodal lines from $2$ in the right dome to $\geq 4$ for the left dome. The band filling at which these two domes are connected is indicated by arrows with the same color as the corresponding dashed lines in  Fig.~\ref{fig:lambda_ipfb}(a). 
When the nematicity is allowed, in a sharp contrast, $\lambda$ becomes significantly higher, and specifically rapidly exceeds unity just below the respective $\ave{n_{\rm c }}$.  After this peak, $\lambda$ 
decreases as the filling is decreased, until a second rise occurs at $\ave{n}=\ave{n_{\rm c_{2}}}$, followed 
by a second decrease.  Corresponding superconducting transition temperatures is displayed  in Fig.~\ref{fig:lambda_ipfb}(b). The double-dome structure is inherited in $T_{\rm c}$, which endorses the interplay between various symmetries of Pomeranchuk distortions and the superconducting instabilities.


 \begin{figure}[t]
    \centering
    \includegraphics[width=0.75\columnwidth]{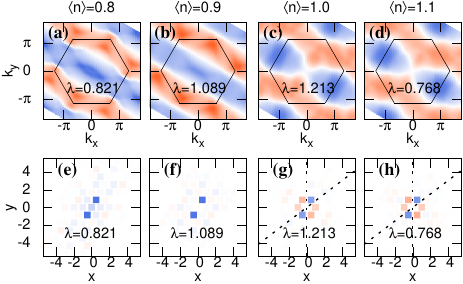}
    \caption{
    Singlet gap functions in momentum-space~(top row) and in real-space~(bottom) for the regular band with $t'=0$ and $\kappa=0$. Black hexagons in top panels indicate the Brillouin zone. Dashed lines in panels (g,h) represent nodes.
    \label{fig:gap_tp0}}
\end{figure}

 \begin{figure}[t]
    \centering
    \includegraphics[width=0.75\columnwidth]{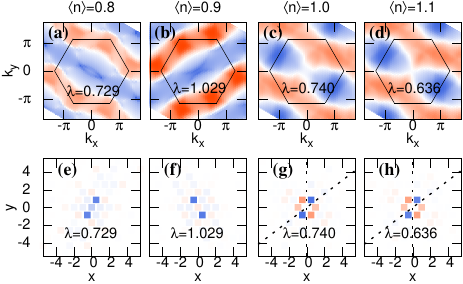}
    \caption{
     Singlet gap functions in momentum-space~(top row) and in real-space~(bottom) for iPFB with $\kappa=1.0$. Black hexagons in top panels indicate the Brillouin zone. Dashed lines in panels (g,h) represent nodes.
    \label{fig:gap_eps1}}
\end{figure}
 \begin{figure}[t]
    \centering
    \includegraphics[width=0.75\columnwidth]{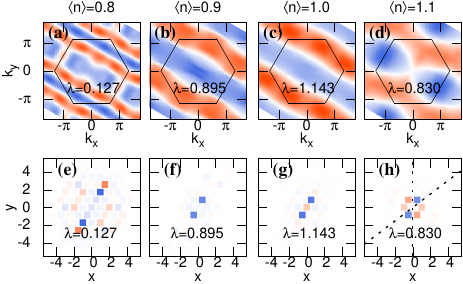}
    \caption{
    The same as above figure for an increased $\kappa=1.5$.
    \label{fig:gap_eps15}}
\end{figure}

To identify the pairing symmetry, we present the singlet gap functions in  momentum  and real spaces for the regular band in Fig.~\ref{fig:gap_tp0}, for iPFB model with $\kappa=1.0$ in Fig.~\ref{fig:gap_eps1}, and for $\kappa=1.5$ in Fig.~\ref{fig:gap_eps15}. Above $\ave{n_{\rm c}}$, the gap functions have $d_{x^2-y^2}$ symmetry~[see panels (d) in these figures], where the pairing in real-space extends to one lattice spacing~[panels (h)].
As in Fig.~4 in the main text for PFB, the gap function changes from the $d_{x^2-y^2}$-wave to a $\gamma s_{x^2+y^2} -d_{x^2-y^2}-d_{xy}$-wave for $\ave{n_{\rm cl}} < \ave{n} < \ave{n_{\rm c}}$ when Pomeranchuk distortions emerge, see panels (a,b) in Fig.~\ref{fig:gap_tp0} for the regular band, panel (a) in Fig.~\ref{fig:gap_eps1} and panels (a,b) in Fig.~\ref{fig:gap_eps15} for iPFB. The value of $\gamma$ increases from zero to almost $0.3$ when the filling is reduced from $ \ave{n_{\rm c}}$ up to $\ave{n_{\rm c_{2}}} $. $\gamma$ acquires values larger than $0.3$ below $\ave{n_{\rm c_{2}}} $. 
In iPFB with $\kappa=1.0$ at $\ave{n}=0.9$ the superconducting gap function exhibits a $\pi/6$ rotated $s_{x^2+y^2} -d_{x^2-y^2}-d_{xy}$-wave pairing, i.e.,$s_{x^2+y^2} -d_{x^2-y^2}+d_{xy}$-wave pairing, which reflects the momentum-dependency of its spin-susceptibility.
 Below $\ave{n_{\rm cl}}$, the pair becomes spatially extended~(panel (e) in Fig.~\ref{fig:gap_eps15}), which accompanies the appearance of multiple nodal lines in the k-space ~(panel (a)). While this is expected for the systems with partially flat bands~\cite{Sayyad2020} because of a bunch of pair-scattering channels around the flat regions, suppression of these types of pairings in the presence of Pomeranchuk instabilities is interesting. This can be understood by recalling that a nematicity, which brings about some superstructures, degrades the flatness of the noninteracting band dispersion. Concomitantly, the spin susceptibility acquires spikes structures.

Aside from the discussed singlet superconductivity, we have also examined the triplet gap functions and associated $\lambda$. Our results show that the $\lambda$ never exceeds $0.4$ in the region studied. This is why we have not presented these data in the present work.

\section{Density of states for triangular lattices}\label{sec:dos}

 \begin{figure}[h]
    \centering
    \includegraphics[width=0.45\columnwidth, height=0.19\textheight]{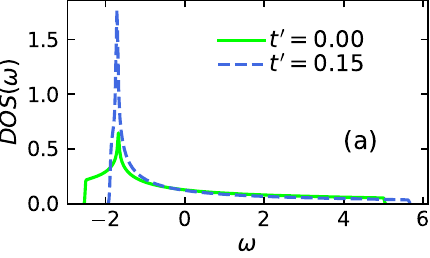}
    \includegraphics[width=0.45\columnwidth, height=0.19\textheight]{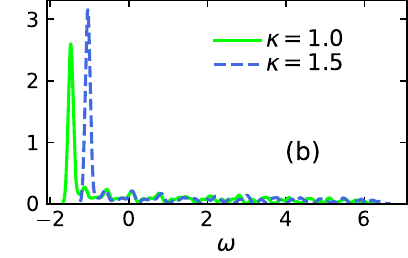}
    \caption{
    Noninteracting density of states in the regular band model with $t'=0$~[solid green line in panel (a)], the PFB model with $t'=0.15$~[dashed blue line in (a)], the iPFB model with $\kappa=1.0$~[solid green line in (b)], or with $\kappa=1.5$~[dashed blue line in (b)]. Note different ranges in the plot between (a) and (b).
    \label{fig:dos}}
\end{figure}

To demonstrate presence of van-Hove singularities in our triangular lattices, we present in Fig.~\ref{fig:dos} the noninteracting density of states~(DOS) for the regular band with $t'=0$~(solid green line in panel (a)), PFB with $t'=0.15$~(dashed blue line in panel (a)), and iPFB with $\kappa=1.0$~(solid green line in panel (b)), or $\kappa=1.5$~(dashed blue line in panel (a)).  The DOS for the system with a partially flat portion, the blue line in panel (a) and lines in panel (b), exhibit a singular value as small energies. A large DOS is also observed at $\omega \approx -1.68$, slightly above the bottom of the band, for the regular band system. Note that while the bandwidths of these four systems are identical, the bottom of their band structure is not located at the same energy.

\section{Further details on the PFB model} \label{app:pfb}

 \begin{figure}[h]
    \centering
    \includegraphics[width=\columnwidth]{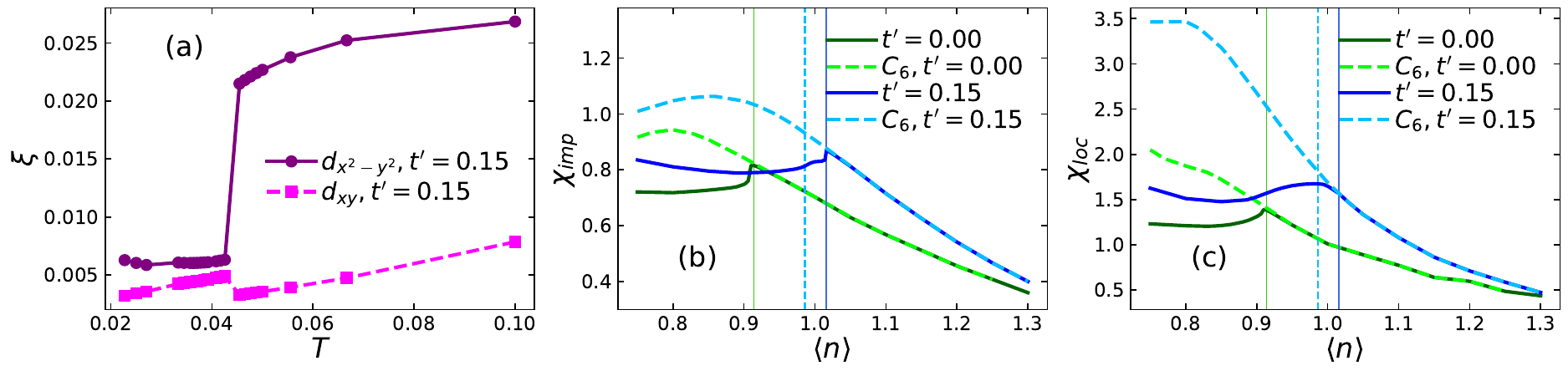}
    \caption{
   (a) For the PFB model Pomeranchuk order parameters, 
   $\xi_{d_{x^2-y^2}}$~(solid purple line) and $\xi_{d_{xy}}$~(pink), at a band filling $\ave{n_{\rm c}}=1.02$. Impurity~(b) and local~(c) spin susceptibilities against band filling are also plotted. All panels are for $U=4.5$ and $\beta=30$. Vertical solid lines in panels (b,c) indicate $\ave{n_{\rm c}}$, while vertical dotted pale-blue lines $\ave{n}=\ave{n_{\rm c_{2}}}$~(see text).
    \label{fig:xi_chiimp}}
\end{figure}

\subsection{Pomeranchuk order parameters at $\ave{n_{\rm c}}$}

To further explore the temperature-dependence of the Pomeranchuk instabilities, we plot $\xi_{d_{x^2-y^2}}$ and $\xi_{d_{xy}}$ at $\ave{n_{\rm c}}$ against temperature in Fig.~\ref{fig:xi_chiimp}(a). While for $T<0.0425$ both of the nematic orders are negligible, $\xi_{d_{x^2-y^2}}$ undergoes a first-order transition at $T=0.0425$, which indicates the broken $C_{6}$ symmetry at higher temperatures.

\subsection{  Impurity spin-susceptibility}

  Impurity spin-susceptibility for PFB model with $t'=0.15$ in Fig.~\ref{fig:xi_chiimp}(b) shows that the increasing spin fluctuations for $\ave{n}> \ave{n_{\rm c}}$ sharply drops when sixfold rotational symmetry is broken for $\ave{n}< \ave{n_{\rm c}}$. Local spin-susceptibility in Fig.~\ref{fig:xi_chiimp}(c), on the other hand, displays a maximum at $\ave{n}=\ave{n_{\rm c_{2}}}$ in PFB.
   We can note that we hardly detect any features at $\ave{n_{\rm c_{2}}}$ in the impurity spin-susceptibility which might be related to a crossover from $\xi_{d_{x^{2}-y^{2}}}$ to $\xi_{d_{xy}}$, see Fig.~\ref{fig:xi_chiimp}(a). When the structure of $\xi$ undergoes an abrupt change from $\xi_{d_{x^{2}-y^{2}}}$ to $\xi_{d_{xy}}$, due to the first-order transition in $\xi_{d_{xy}}$, the impurity spin-susceptibility exhibits a kink at $\ave{n_{\rm c_{2}}}$, see e.g.  Fig.~\ref{fig:docc_ipfb}(b) and related discussion in Sec.~\ref{app:ipfb}.

\section{Dependency of the critical band filling on the Hubbard interaction} \label{app:docc}

 \begin{figure}[h]
    \centering
    \includegraphics[width=0.8\columnwidth]{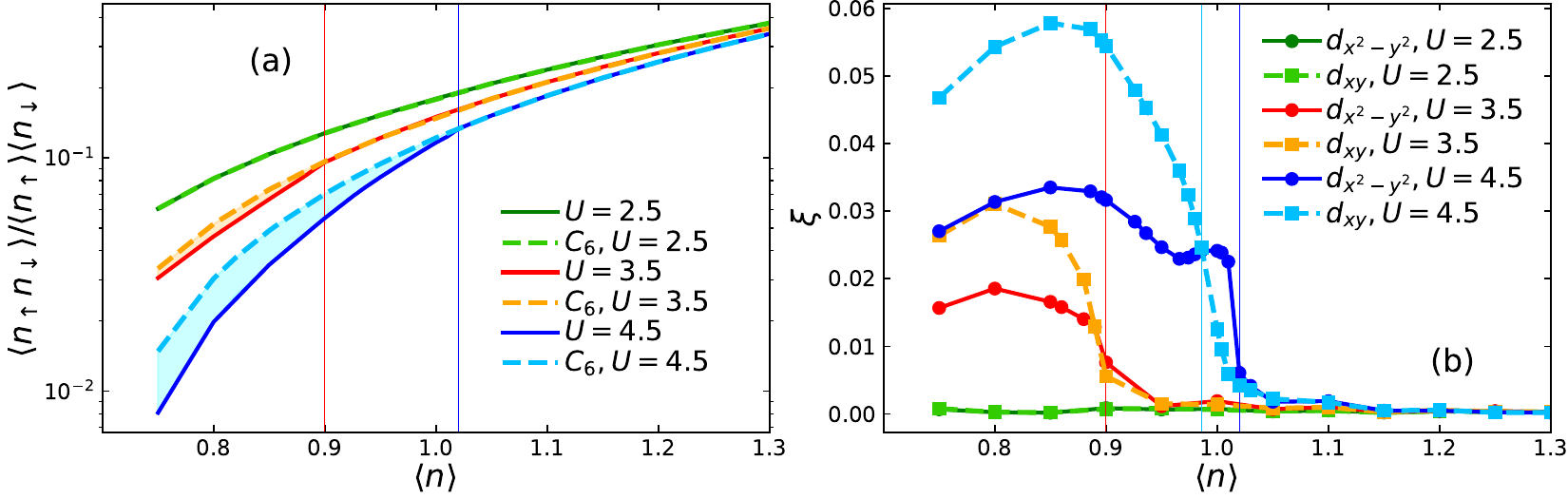}
    \caption{
    Normalized double occupancy~(a) and Pomeranchuk order parameter~(b) for the PFB systems with $t'=0.15$, $U=$ $2.5$~(green lines), $3.5$~(red lines), and $4.5$~(blue lines) and at $\beta=30$. Dashed lines in panel (a) describe normalized double occupancies for systems with the imposed sixfold symmetry. $\xi_{d_{x^2-y^2}}$ and $\xi_{d_{xy}}$ are presented with solid and dashed lines in panel~(b), respectively. Vertical lines indicate $\ave{n_{\rm c}}$.
    \label{fig:docc_3U}}
\end{figure}

To explore the relation between $\ave{n_{\rm c}}$ and the Hubbard interaction, we present normalized double occupancies and Pomeranchuk order parameters for three different $U$ values in Fig.~\ref{fig:docc_3U}. The sixfold symmetry of the triangular lattice remains unbroken for the PFB systems with $t'=0.15$ and $U=2.5$ as can be seen both from the renormalized double occupancy, green lines in panel~(a), and also from zero Pomeranchuk instabilities, green lines in panel~(b).
The $C_{6}$ symmetry is broken at larger Hubbard interactions, namely $U=\{ 3.5,4.5 \}$, as shown by red and blue lines in Fig.~\ref{fig:docc_3U}, respectively. The critical band filling $\ave{n_{\rm c}}$ for the PFB systems with $U=3.5$ is $0.9$~(red vertical lines).   

\subsection{Superconductivity}

  \begin{figure}[t]
     \centering
     \includegraphics[width=0.5\columnwidth]{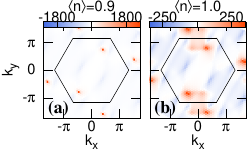}
     \caption{
     Difference, $\Delta V_{\rm eff} = V_{\rm eff}- V_{\rm eff}^{C_{6}}$, in the effective pairing interaction between the absence and presence of the $C_{6}$ symmetry for the PFB system with $t'=0.15$ for $U=4.5$ and $\beta=30$ for $\ave{n}=0.9$ (a) and 1.0(b). Black hexagons represent the Brillouin zone. 
     Note that the scale of the color bar differs by orders of magnitude between the panels.
     \label{fig:diff_Veff}}
 \end{figure}
 
Fig.~\ref{fig:diff_Veff} presents $\Delta V_{\rm eff} = V_{\rm eff}- V_{\rm eff}^{C_{6}}$, where $V_{\rm eff}^{C_{6}}$ is the effective interaction with the imposed $C_{6}$ constraint. We can see that $\Delta V_{\rm eff}$ displays not only the broken $C_{6}$ symmetry but also amplitudes significantly intensified than in the $C_{6}$ case. See also the discussion on $\Delta V_{\rm eff}$ in the main text.

 \begin{figure}[t]
    \centering
    \includegraphics[width=0.5\columnwidth]{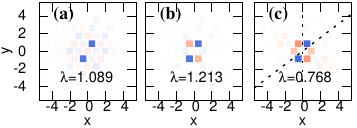}
    \caption{
    Gap functions in real space with 
    singlet pairing for the PFB system with $t'=0.15$ for $U=4.5$ and $\beta=30$ for $\ave{n}=0.9$ (a), 1.0(b) and 1.1(c). Dashed lines in panel (c) represent nodes. Color code for the gap function is bluish~(reddish) for negative~(positive)
values, for which we have omitted the color bars since the
linearized Eliashberg equation does not indicate magnitudes
of $\Delta$. This figure is to be compared with Fig.~4(b-d) in the momentum space in the main text.
    \label{fig:gap_r_app}}
\end{figure}

Gap functions in the real-space with singlet, even-frequency pairing for the PFB model with $t'=0.15$ is plotted in Fig.~\ref{fig:gap_r_app}. In the main text, we presented the $\gamma s -d_{xy}$-wave symmetry of the gap function below $\ave{n_{\rm c}}$ in the momentum space; see Fig.~4(b-d). Evidently, the electron pairing is short-range in this filling region. This observation is in contrast with the formation of extended Cooper pairs in systems where the $C_{6}$ constraint is imposed.

\section{Possible singlet pairing symmetries on the triangular lattice}~\label{appsec:gap}

 \begin{figure}[h]
    \centering
    \includegraphics[width=0.65\columnwidth]{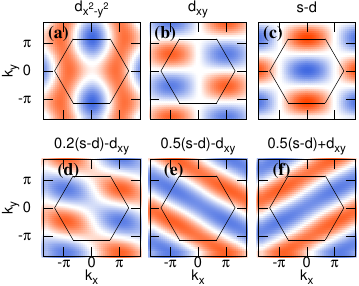}
    \caption{
    Gap functions $d_{x^2-y^2}$~(a), $d_{xy}$~(b), $s_{x^2+y^2} -d_{x^2-y^2}$~(c), $0.2(s_{x^2+y^2} -d_{x^2-y^2})-d_{xy}$~(d), $0.5(s_{x^2+y^2} -d_{x^2-y^2})-d_{xy}$~(e) and $0.5(s_{x^2+y^2} -d_{x^2-y^2})+d_{xy}$~(f) 
    are plotted in momentum-space, with positive~(negative) values represented by 
    blue~(red). $s-d$ in panels (c-f) is the shorthand for $s_{x^2+y^2} -d_{x^2-y^2}$.
    \label{fig:gapfuncs}}
\end{figure}

In this section, we present various symmetries in the gap functions that can be realized on the triangular lattice. In the presence of $C_{6}$ symmetry and 
for the pairs not extending beyond nearest neighbors in real space, the spin-singlet gap functions are either $s_{x^2 +y^2}$-wave or $d$-wave, with respective forms in k-space,
\begin{align}
\Delta_{s_{x^2 +y^2}}(\bolds{k}) &= \cos(k_x) + 2 \cos(\sqrt{3}k_y/2) \cos(k_x/2) ,\\
    \Delta_{d_{x^2 -y^2}}(\bolds{k}) &= \cos(k_x) - \cos(\sqrt{3} k_y/2) \cos(k_x/2) ,\\
    \Delta_{d_{xy}}(\bolds{k}) &=
    \sqrt{3} \sin( \sqrt{3}k_y/2) \sin(k_x/2).
\end{align}
In the presence of $C_{6}$ symmetry, the above two $d$-waves, depicted in Fig.~\ref{fig:gapfuncs}(a-b), are associated with two degenerate irreducible representations in the $C_{6}$ point group, so that the topological superconductivity with gap function $\Delta_{d_{x^2 -y^2}}(\bolds{k}) + \i \Delta_{d_{xy}}$ is also possible.
When $C_{6}$ symmetry is reduced  down to $C_{2}$, on the other hand, $\Delta_{s_{x^2 +y^2}}$ and $\Delta_{d_{x^2 -y^2}}$ are no longer irreducible representations, and instead their linear combination $s_{x^2+y^2} -d_{x^2-y^2}$, $\Delta_{s-d}$, with a form factor,
\begin{align}
    \Delta_{s-d}(\bolds{k}) &=  3 \cos(\sqrt{3}k_y/2) \cos(k_x/2) ,
\end{align}
finds room to emerge, see Fig.~\ref{fig:gapfuncs}(c). In the $C_{2}$ point group, 
$s_{x^2+y^2} -d_{x^2-y^2}$ reads $s_{y2}$, and in this point 
group $s_{y2}$ and $d_{xy}$ are not degenerate, so that this group does not permit topological superconductivity. Still, linear combinations of $\Delta_{d_{xy}}$ and $\Delta_{s-d}$ may be realized, as we have indeed shown in Figs.~4(bottom rows), \ref{fig:gap_tp0}, \ref{fig:gap_eps1}, \ref{fig:gap_eps15}. Fig.~\ref{fig:gapfuncs}(d-f) shows $\alpha (s_{x^2+y^2} -d_{x^2-y^2})+ \beta d_{xy}$ with $(\alpha, \beta) =$ $(0.2,-1)$ in (d), $(0.5,-1)$ in (e) and $(0.5,+1)$ in (f). Panel (d) in Fig.~\ref{fig:gapfuncs} exhibits a momentum-dependence similar to Fig.~4(c) on the PFB model at half filling. So this is why we can say the pairing in the PFB model is $s_{x^2+y^2} -d_{x^2-y^2} - d_{xy}$.
The structure exemplified by panel (e) is evident in most gap functions below $\ave{n_{c_{2}}}$ in the regular band, PFB and iPFB models. The gap function in panel (f) is also displayed at $\ave{n}=0.9$ in the iPFB model with $\kappa=1.0$ in Fig.~\ref{fig:gap_eps1}(b).

\bibliography{flat.bib}

\begin{thebibliography}{51}
\bibitem [{eigp()}]{eigp}%
  \BibitemOpen
  \bibinfo {note} {We have checked that the values of $\lambda$ for the
  spin-triplet even-frequency, singlet and triplet odd-frequency gap functions
  are much smaller than those for the spin-singlet even-frequency.}\BibitemShut
  {Stop}%
\bibitem [{Extendpair()}]{Extendpair}%
  \BibitemOpen
  \bibinfo {note} {In SM, we present spatially extended pairings, with
  multi-nodal gap functions for $\ave{n}<\ave{n}_{\rm min}$~(marked with a blue
  arrow in Fig.~3) in PFB with $C_{6}$ symmetry~\cite{Suppmat}. However, such
  pairings do not form when nematicity is present. As a result, the weight of
  the real-space gap function in PFB mainly extend to the nearest
  neighbors.}\BibitemShut {Stop}%
\end{thebibliography}

\end{document}